\documentstyle[epsf]{article}          
\def\ennu{\left(\rho_\nu + \rho_{\bar\nu} \right)_{_0}}
\def\nnu{\left(n_\nu - n_{\bar\nu} \right)_{_0}}

\title{\begin{flushright}
\small August 1998 \hfill SINP/TNP/98-21\\ 
\tt hep-ph/9809410
\end{flushright}
\bf How degenerate can cosmological neutrinos be?}

\author{
{\bf 
Palash B. Pal\thanks{Email: pbpal@tnp.saha.ernet.in}
~{\rm and}\ 
Kamales Kar\thanks{Email: kamales@tnp.saha.ernet.in}
}
\\ 
\normalsize 
Saha Institute of Nuclear Physics, 1/AF Bidhan-Nagar, 
Calcutta 700064, India}

\date{}

\begin{document}

\maketitle

\begin{abstract}
\noindent
There are well-known bounds on light neutrino masses from cosmological
energy density arguments. These arguments assume the neutrinos to be
non-degenerate. We show how these bounds are affected if the neutrinos
are degenerate. In this case, we obtain correlated bounds between
neutrino mass and degeneracy.
\bigskip\bigskip
\end{abstract}

In the standard big bang cosmology, neutrinos play a very important
role in the evolution of the universe \cite{MohPal}. As a result,
neutrino properties can be significantly constrained from various
cosmological parameters. Perhaps the most famous of such constraints
is the one on the mass of light neutrinos \cite{GeZe66}. Using modern
values of the cosmological parameters some of which we will discuss
later, this bound comes out to be
	\begin{eqnarray}
\sum_i m_i < 46 \; {\rm eV}\,,
\label{46}
	\end{eqnarray}
where $m_i$ is the mass of the light neutrino $\nu_i$.

The derivation of this bounds assumes that the cosmological neutrinos
are non-degnerate, i.e., the number of neutrinos and antineutrinos are
equal. If this condition is not satisfied, the bound is modified. For
example, if the neutrinos are assumed to be massless and at zero
temperature, the degenerate Fermi gas of neutrinos can oversaturate
the energy density of the universe unless the chemical potentials
$\mu_i$ associated with the neutrinos $\nu_i$ satisfy the
constraint~\cite{Weinberg}
	\begin{eqnarray}
\left[ \sum_i \mu_i^4 \right]^{1/4} < 7.4 \times 10^{-3} \; {\rm eV}\,.
\label{mubound}
	\end{eqnarray}

The two cases discussed above are extreme cases --- one a bound on
mass for vanishing chemical potential, and the other a bound on the
chemical potential for vanishing mass. More generally, both mass and
chemical potential may be non-zero. One then should obtain correlated
bounds on mass and degeneracy of the neutrinos. In view of the
importance of the neutrinos in the physics of the early universe, such
bounds are potentially important. The purpose of this paper is to
derive such bounds. For the sake of simplicity, we assume that only
one light neutrino species dominates the energy density of the
universe.

To discuss these bounds, it is useful to introduce the parameter
$F_\nu$, defined by
	\begin{eqnarray}
F_\nu = {\ennu \over \rho_0} \,,
\label{Omnu}
	\end{eqnarray}
where the numerator is the sum of the present energy densities of the
neutrinos and antineutrinos, and $\rho_0$ is the total energy density
of the present universe. Obviously, $F_\nu<1$. Modern theories of
structure formation in the universe prefer \cite{structure} a value
around $0.25$ for this parameter. We will present our results for
various values of this parameter.

The total energy density $\rho_0$ is not very well-known. It is
usually parametrized in the form
	\begin{eqnarray}
\rho_0 = 10^4 h^2 \Omega_0 \; {\rm eV/cm}^3 \,,
\label{rho0}
	\end{eqnarray}
where $\Omega_0$ is the density measured in units of the critical
density, the latter being dependent on the Hubble parameter, which in
turn is parametrized as $100h\;{\rm km\,s^{-1}\,Mpc^{-1}}$. Combining
Eqs.\ (\ref{Omnu}) and (\ref{rho0}), we can write
	\begin{eqnarray}
\ennu = 10^4 \zeta_\nu \; {\rm eV/cm}^3 \,,
\label{rhonu}
	\end{eqnarray}
where
	\begin{eqnarray}
\zeta_\nu \equiv h^2 \Omega_0 F_\nu \,.
\label{zetanu}
	\end{eqnarray}
In the rest of the paper, we will take different values for this
parameter $\zeta_\nu$ which measures the importance of the neutrinos
in the early universe, and will find the combinations of neutrino mass
and degeneracy which can produce the corresponding values for
$\ennu$. The bounds quoted in Eqs.\ (\ref{46}) and (\ref{mubound})
correspond to $\zeta_\nu=0.5$.

At any epoch, the energy density of neutrinos plus antineutrinos can
be written as
	\begin{eqnarray}
\rho_\nu + \rho_{\bar\nu} &=& \int {d^3p \over (2\pi)^3} \; \sqrt
{p^2+m^2} \left[ 
f(p) + \bar f(p) \right] \nonumber\\*
&=& {1\over 2\pi^2} \int_0^\infty dp \; p^2  \sqrt {p^2+m^2} \left[
f(p) + \bar f(p) \right] \,,
\label{genrho}
	\end{eqnarray}
where $f(p)$ and $\bar f(p)$ denote the distribution functions of the
neutrinos and the antineutrinos at that epoch, $p$ being the
magnitude of the momentum 3-vector.

\begin{figure}
\centerline{\epsfxsize=0.6\textwidth
\epsfbox{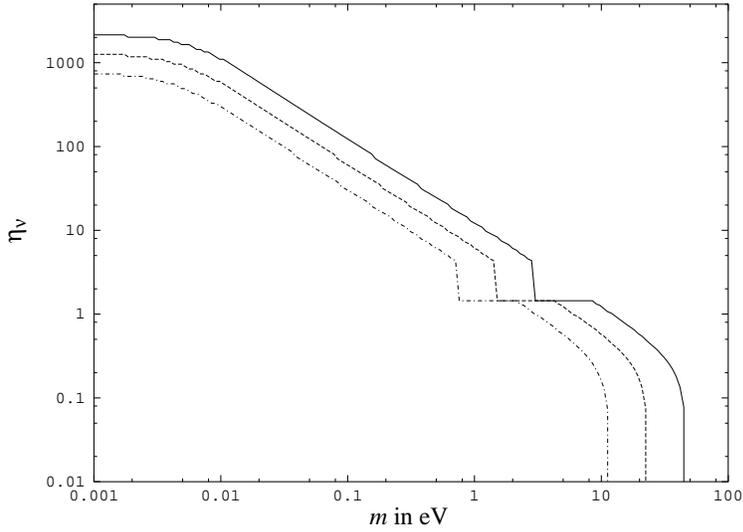}}
\caption[]{\sf The contours of equal energy density of neutrinos plus
antineutrinos as a function of neutrino mass and the degeneracy
parameter $\eta_\nu$ introduced in Eq.\ (\ref{etanu}). From the outer
to the inner curve, the values of $\zeta_\nu$ are $1\over2$, $1\over4$
and $1\over8$ respectively. The tick marks on the axes are at 2, 5 and
8 of each decade.}\label{f.bounds}
\end{figure}

In the very early universe, the neutrinos were in equilibrium with
other particles, so $f(p)$ and $\bar f(p)$ were the Fermi-Dirac
distribution functions, with appropriate temperature and chemical
potential. Due to the expansion of the universe, there reached an
epoch when the neutrino interactions were not enough to keep them in
thermal equilibrium. This is called the epoch of neutrino
decoupling. At this time, let the temperature of the neutrinos be
$T_D$ and the chemical potential $\mu_D$. Then, at this epoch, the
energy density of neutrinos and antineutrinos was given by Eq.\
(\ref{genrho}), with
	\begin{eqnarray}
f_D (p) = {1 
\over \exp \left( {\sqrt{p^2+m^2} - \mu_D \over T_D} \right) +1} \,,
\label{fd}
	\end{eqnarray}
and that of the antineutrinos can be obtained by changing the sign of
$\mu_D$. After that time, the neutrino momenta did not change at all
due to collisions. They only suffered the cosmological red-shift. If
the cosmological scale parameter was $a_D$ at the decoupling era and
$a_0$ now, all the momenta have decreased by a factor
	\begin{eqnarray}
r \equiv {a_D\over a_0} \,.
\label{r}
	\end{eqnarray}
The quantity $r$ can be interpreted as the ratio of the scale factors
of the era when the neutrinos were last in thermal equilibrium, and
the present era. The distribution function of the neutrinos in the
present universe can be obtained from Eq.\ (\ref{fd}) through the
relation
	\begin{eqnarray}
f_0(p) = f_D(p/r) \,.
\label{f0}
	\end{eqnarray}
If $m=\mu=0$, this is still a thermal distribution with a redefined
temperature $rT_D$. When either the mass or the chemical potential is
not zero, the distribution is {\em not}\/ thermal. Making a simple
change of variable, we can write
	\begin{eqnarray}
\ennu = {r^3\over 2\pi^2} 
\int_0^\infty dp \; p^2  \sqrt {r^2 p^2+m^2} \left[
f_D(p) + \bar f_D(p) \right] \,.
\label{ennu}
	\end{eqnarray}
Since the distribution is not thermal as we have already said, it is
meaningless to talk about the chemical potential of neutrinos in the
present universe. Rather, we can talk about the difference of the
number densities of neutrinos and antineutrinos. Using similar
argument, it will be given by
	\begin{eqnarray}
\nnu = {r^3\over 2\pi^2} 
\int_0^\infty dp \; p^2  \left[
f_D(p) - \bar f_D(p) \right] \,.
\label{nnu}
	\end{eqnarray}
In Fig.~\ref{f.bounds}, we have shown for a range of neutrino mass
$m$, the values of the degeneracy parameter
	\begin{eqnarray}
\eta_\nu \equiv \nnu / n_\gamma 
\label{etanu}
	\end{eqnarray}
which give certain preassigned values for the density paramter
$\zeta_\nu$. In Eq.\ (\ref{etanu}), $n_\gamma$ is the number density
of photons in the microwave background which has been introduced to
obtain a dimensionless parameter for neutrino degeneracy. We will
discuss the characteristics of these results later. Before that, we
want to discuss different considerations that go into obtaining the
results. 

{}From the previous discussion, it seems that if we want to find out
the energy density for the neutrinos, we need to know $m$, $\mu_D$,
$T_D$ and $r$. The plot of Fig.~\ref{f.bounds} is two dimensional, so
we need to show that two of these variables are dependent. We take $m$
and $\mu_D$ as independent. Given these two, we need to solve for the
temperature at which the reaction rate of the neutrinos become equal
to the expansion rate of the universe. For this part, we can use $r=1$
because, until decoupling took place, the neutrinos were in thermal
equilibrium. Once $T_D$ is obtained this way, we can put
	\begin{eqnarray}
r = y_{\rm rh}\; {T_0\over T_D} 
\label{rT}
	\end{eqnarray}
for the calculation of energy and number densities in the present
universe, where $T_0$ is the temperature associated with the microwave
background. For $y_{\rm rh}=1$, this follows from Eq.\ (\ref{f0}) and
the comment following it, which is valid for a Bose-Einstein
distribution as well. The issue of the departure of $y_{\rm rh}$ from
the value of unity will be discussed next.

The value of $y_{\rm rh}$ should be equal to unity provided the number
of photons have not increased due to the annihilation of other species
of particles after the neutrinos decoupled. Thus, if the neutrino
decoupling temperature is smaller than the electron mass, we should
put $y_{\rm rh}=1$. If, on the other hand, $T_D>m_e$, we should take
into account the reheating of photons from $e^-$-$e^+$
annihilations. This gives \cite{MohPal} $y_{\rm
rh}=(4/11)^{1/3}$. These considerations have been taken into account
in the plot of Fig.~\ref{f.bounds}, as we discuss below.

The decoupling temperature is determined by using the criterion
$\Gamma=H$, where $\Gamma$ denotes the reaction rate for neutrinos and
$H$ is the Hubble parameter. In the early universe, $H$ is given by
$T^2/M_P$, apart from numerical factors of order unity. On the other
hand, neutrino reaction rates are given by $\Gamma=n\sigma$, where $n$
is the number density of the particles that neutrinos interact with,
and $\sigma$ is the scattering cross section. When neutrinos are
highly degenerate, they are more abundant than electrons and
positrons. Therefore, elastic scattering as well as $\nu$-$\bar\nu$
annihilation are the most efficient mechanisms for redistribution of
energy and momentum of neutrinos, which keep them in thermal
equilibrium. Hence, in the expression for the reaction rate, one
should use $n=n_\nu+n_{\bar\nu}$. The cross section is given by
$G_F^2E^2$ where $E$ is the typical energy of the neutrinos. Since the
masses are small in the range under consideration, we can approximate
$E$ by the temperature $T$. Therefore, the decoupling temperature is
determined from the equation
\begin{eqnarray}
n_\nu+n_{\bar\nu} = {1\over G_F^2 M_P} \,,
\end{eqnarray}
where $n_\nu$ and $n_{\bar\nu}$ should be determined using the
equilibrium distribution functions.

The plots have been made for three different values for the density of
the neutrinos and antineutrinos. First, considerations of the age of
the universe dictates \cite{age} that $h^2\Omega_0$ cannot be larger
than about 0.5. This implies that $\zeta_\nu\leq 0.5$. We make the
plots corresponding to three values of $\zeta_\nu$, viz., 1/2, 1/4 and
1/8. If $h^2\Omega_0$ is indeed 1/2, these values of $\zeta_\nu$
corresponds to neutrinos and antineutrinos contributing 100\%, 50\%
and 25\% respectively to the energy density of the universe.

The plots for a given density clearly shows three regions. First,
there is a vertical region for vanishingly small degeneracy. Here, our
bound agrees exactly with Eq.\ (\ref{46}), as it should. In fact, the
outer vertical line has an intercept on the horizontal axis exactly at
46 eV. Second, for small or vanishing mass, we obtain a horizontal
branch which corresponds to the bound of Eq.\ (\ref{mubound}). Here,
the intercept on the vertical axis is $2.2\times10^3$, which is the
number obtained from Eq.\ (\ref{mubound}). Finally, there is an
intermediate region which represents the new results obtained in this
paper. This part follows roughly a power law behavior, given by
\begin{eqnarray}
\eta_\nu \left( {m \over 1\,{\rm eV}} \right) = 24 \zeta_\nu \,.
\end{eqnarray}
The glitch encountered in this region corresponds to the shift of the
value of $y_{\rm rh}$ from $(4/11)^{1/3}$ to unity. Earlier, we
explained how we have calculated the decoupling temperature $T_D$. If
$T_D$ is less than the electron mass, we have taken $y_{\rm rh}$ to be
unity. Otherwise, it is taken to be $(4/11)^{1/3}$. Because of these
sudden approximations, the shift appears as a glitch. In a more
detailed calculation of the densities of different particles,
involving the Boltzmann equation, there should be a more gradual shift
from one value of $y_{\rm rh}$ to another.

The usefulness of these plots is the following. There are quite a few
recent indications that neutrinos are massive. These include the solar
neutrino data \cite{solar}, the atmospheric neutrino data \cite{atmos}
and direct neutrino oscillation expermients \cite{oscill}. Each kind
of data points to a different range of neutrino mass, but none of them
very close to the value obtained from cosmological dark matter
considerations assuming zero degeneracy for neutrinos and taking
values of $h^2\Omega_0$ close to $1/2$.  The point that we make here
is that, if we introduce neutrino degeneracy as a free parameter, we
can obtain solutions for any energy density given the value of
neutrino mass. Such a solution would of course be acceptable if it
conforms with other bounds on neutrino degneracy, e.g., those coming
from primordial nucleosynthesis \cite{subir}.  One can also ask
whether, given a certain value of mass, the required value of neutrino
degeneracy is a plausible one. This can only be addressed within a
certain scenario of generation of lepton asymmetry in the
universe. This is a separate issue and should be taken up separately.

\paragraph*{Note added~: } 
After this paper was accepted for publication, we came to know about a
paper by Khare and Deo \cite{khare} which also found the correlated
bounds on neutrino mass and degeneracy from the upper limit on
cosmological energy density. However, these authors assumed the
neutrino distribution to be thermal even in the present universe. In
our analysis, we have taken into account neutrino decoupling and the
subsequent deviation of the neutrino distribution from a thermal one.


\end{document}